\documentclass[12pt]{article}

\usepackage{scicite}

\usepackage{times}

\newenvironment{sciabstract}{%
\begin{quote} \bf}
{\end{quote}}

\newcounter{lastnote}

\title{Suppressing Roughness of Virtual Times in Parallel
Discrete-Event Simulations}

\author
{G. Korniss,$^{1\ast}$ M.A. Novotny,$^{2}$ H. Guclu,$^{1}$
Z. Toroczkai,$^{3}$ P.A. Rikvold$^{4}$\\
\\
\normalsize{$^{1}$Department of Physics, Applied Physics, and
Astronomy, Rensselaer Polytechnic Institute,}\\
\normalsize{110 8$^{th}$ Street, Troy, NY 12180-3590, USA.}\\
\normalsize{$^{2}$Department of Physics and Astronomy and ERC Center
for Computational Science,}\\ 
\normalsize{Mississippi State University, P.O.\ Box 5167, Mississippi
State, MS 39762-5167, USA.}\\ 
\normalsize{$^{3}$Complex Systems Group, Theoretical Division,
Los Alamos National Laboratory,}\\
\normalsize{MS B-213 Los Alamos, NM 87545, USA.}\\
\normalsize{$^{4}$Department of Physics, Center for Materials Research and 
Technology,}\\ 
\normalsize{and School of Computational Science and Information Technology,}\\
\normalsize{Florida State University, Tallahassee, FL 32306-4350, USA.}\\
\\
\normalsize{$^\ast$To whom correspondence should be addressed; E-mail:
korniss@rpi.edu.} 
}

\date{}  %Include the date command, but leave its argument blank.

\begin{document} 

\baselineskip24pt  % Double-space the manuscript.

\maketitle 

\enlargethispage*{1000pt}

\begin{sciabstract}
In a parallel discrete-event simulation (PDES) scheme, tasks are distributed
among processing elements (PEs), whose progress is controlled by a
synchronization scheme. For lattice systems with short-range interactions,
the progress of the conservative PDES scheme is governed by the
Kardar-Parisi-Zhang equation from the theory of non-equilibrium
surface growth. Although the simulated (virtual) times of the PEs progress at
a nonzero rate, their standard deviation (spread) diverges
with the number of PEs, hindering efficient data collection. We
show that weak random interactions among the PEs can make
this spread nondivergent. The PEs then progress at a nonzero,
near-uniform rate without requiring global synchronizations.
\end{sciabstract}

\thispagestyle{empty}

%\noindent
{Science {\bf 299}, 677 (2003); 
http://www.sciencemag.org/cgi/reprint/299/5607/677.pdf

\pagebreak

%%%%%%%%%%% MAIN TEXT %%%%%%%%%%%%%%%%%%%%%%%%%%%%%%%%%%%%%%%%%%%%%

Simulating large systems often leaves the programmer with only 
one option: parallel distributed simulations where parts of the
system are allocated and simulated on different processing
elements (PE). A large class of interacting systems, including
financial market models, epidemic models, dynamics of magnetic
systems, and queuing networks, can be described by a set of local
state variables assuming a finite number of possible values. As
the system evolves in time, the values of the local state
variables change at discrete instants, synchronously or
asynchronously depending on the dynamics of the system. 
Parallel simulation for the former is straightforward (at least
conceptually). For the latter, i.e., for
asynchronous or non-parallel dynamics, one must 
use some kind of synchronization to ensure causality. The
instantaneous changes in the local configuration are also called
discrete events, hence the term parallel discrete-event
simulations (PDES) \cite{FUJI90,NICOL94,LUBA00}. Examples of PDES applications
include dynamic channel allocation in cell-phone communication
networks \cite{LUBA00,GREEN94}, models the spread of diseases
\cite{CARACO96}, battle-field simulations \cite{NICOL88}, 
and dynamic phenomena in highly anisotropic
magnetic systems \cite{KORN99_JCP,KORN01_PRE}. Here the discrete
events are call arrivals, infections, troop movements, and changes of
the orientation of the local magnetic moments, respectively. As
the number of PEs on parallel architectures
increases to tens of thousands, fundamental questions of the
scalability of the underlying algorithms must be addressed. Here
we show a way to construct fully scalable parallel
simulations for systems with asynchronous dynamics and
short-range interactions. Understanding the effects of the
microscopic dynamics (corresponding to the algorithmic
synchronization rules) on the global properties of the 
simulation scheme brings us to the solution. Recently, a similar
connection has been made \cite{SLOOT} between rollback-based PDES schemes
\cite{JEFF85} and self-organized criticality \cite{BAK87}.

The two basic ingredients of PDES are the set of local simulated
times, often referred to as virtual times \cite{JEFF85}, and
a synchronization scheme \cite{FUJI90}. 
In order for the PDES scheme to be scalable \cite{GREEN96},
({\it i}) the virtual time horizon should progress on average at
a nonzero rate; ({\it ii}) the typical spread of the time horizon
should be bounded as the number of PEs $N_{\rm PE}$ goes to
infinity. The first criterion ensures a nonzero progress rate in the
limit of large $N_{\rm PE}$. It is,
however, not sufficient if data are to be collected. Different PEs
have progressed to different local simulated times with a possibly
large spread among them, making measurement a complex task. 
Frequent global synchronizations can get costly for large $N_{\rm PE}$,
whereas temporarily storing a large amount of data as a result of
the large virtual time spread is limited by the available
memory. Therefore, ({\it ii}) is crucial for the 
measurement part of the algorithm to be scalable. Here we introduce a
PDES scheme in which the PEs make nonzero and close-to-uniform progress
without global intervention.

In conservative PDES schemes \cite{CHAND81,LUBA87,LUBA88}, which we
focus on, an update is performed by a particular PE only if the resulting
change in the local configuration of the simulated system is guaranteed
not to violate causality. Otherwise the PE
idles. The efficiency of the scheme depends on the fraction of
non-idling PEs. It was shown \cite{KORN00_PRL,KORN00_UGA} that the
virtual time horizon exhibits kinetic roughening \cite{BARA95,ZHANG95}
for the basic conservative scheme applied to systems with
short-range interactions on regular lattices.
In particular, the evolution of the virtual time
horizon is governed by the Kardar-Parisi-Zhang (KPZ) equation
\cite{KARD86}, which plays a central role in non-equilibrium
surface growth \cite{BARA95,ZHANG95}. The above finding has
two major implications for the asymptotic scalability of the basic
conservative PDES scheme \cite{KORN00_PRL,TORO00}: criterion ({\it i}) for
the scalability is satisfied because the average progress rate of the
virtual time horizon approaches a nonzero value in the limit 
$N_{\rm PE}\to \infty$.  Criterion ({\it ii}), however, is violated because
the virtual time horizon becomes macroscopically rough.

For illustration, we consider a general one-dimensional system
with nearest-neighbor interactions, in which the discrete events
exhibit Poisson asynchrony. In the one-site-per PE scenario, each
site has its own local simulated time, constituting the
virtual time horizon $\{\tau_{i}(t)\}_{i=1}^{N_{\rm PE}}$. 
Here $t$ is the discrete number of parallel steps executed by
all PEs, which is proportional to the wall-clock time. According to the basic
conservative synchronization scheme \cite{LUBA87,LUBA88}, at each
parallel step $t$, only those PEs for which the local simulated
time is not greater then the local simulated times of their
neighbors, can increment their local time by an exponentially
distributed random amount. (Without loss of generality we assume
that the mean of the local time increment is one in simulated time
units [stu].) Thus, if
$\tau_{i}(t)\leq\min\{\tau_{i-1}(t),\tau_{i+1}(t)\}$, PE
$i$ can update the configuration of the underlying site it
carries and determine the time of the next event. Otherwise, it
idles. Despite its simplicity, this rule preserves unaltered the asynchronous
causal dynamics of the underlying system \cite{LUBA87,LUBA88}. 
Further, the progress rate of the simulation $\langle
u(t)\rangle_{N_{\rm PE}}$ (the density of local minima of the virtual
time horizon) approaches a nonzero constant in the
asymptotic long-time, large-$N_{\rm PE}$ limit
\cite{KORN00_PRL,TORO00}. The average width of the virtual time
horizon, however, diverges as $N_{\rm PE}\!\to\!\infty$
\cite{KORN00_PRL,KORN00_UGA}. Specifically, the average width is defined as
%%%%%%%%%%%%%%%%%%%%%%%%%%%%%%%%%%%%%%%%%%%%%%%%%%%%%%%%%%%%%%%%%%%%
$\left\langle w^2(t)\right\rangle_{N_{\rm PE}}\!=\!(1/N_{\rm PE})
\left\langle \sum_{i=1}^{N_{\rm
PE}}\left[\tau_i(t)-\overline\tau(t)\right]^2 \right\rangle $
%%%%%%%%%%%%%%%%%%%%%%%%%%%%%%%%%%%%%%%%%%%%%%%%%%%%%%%%%%%%%%%%%%%%
where $\overline\tau(t)\!=\!(1/N_{\rm PE})\sum_{i=1}^{N_{\rm
PE}}\tau_i(t)$. For any finite number of PEs, the width grows for
early times as $\left\langle
w^2(t)\right\rangle_{N_{\rm PE}}\!\sim\!t^{2\beta}$, and after a system-size
dependent cross-over time $t_{\times}\!\sim\!N_{\rm PE}^{z}$
it reaches its steady state with $\left\langle
w^2(\infty)\right\rangle_{N_{\rm PE}}\!\sim\!N_{\rm
PE}^{2\alpha}$. Here $\alpha$ is
the roughness exponent (equal to $1/2$ in our example), which quantifies how
the average width of the surface diverges in the large-$N_{\rm PE}$
limit. The reason for this divergence is that
the local interaction topology of the PEs mimicks that of the
underlying system. Under the basic conservative synchronization
rules, the PEs form a strongly interacting system in which the
correlation length reaches the system size $N_{\rm PE}$
\cite{KORN00_PRL}.  This correlation length is responsible for the
diverging surface fluctuations captured by the average width.

To reduce the width, one must de-correlate the fluctuations of
the virtual time surface. We achieve this by introducing quenched
random communication links between the PEs in addition to
the regular nearest-neighbor interactions of the
underlying physical system. At the beginning of the simulation we
connect each PE to another one, chosen randomly from the
rest. Note that these random links in addition to the regular lattice
connections result in a small-world network
\cite{WATTS}, where beyond the nearest neighbors, each PE
is connected with a randomly chosen one \cite{NEWMAN}. In the modified
conservative PDES scheme, at every parallel step each PE with
probability $p$ compares its local simulated time with its full
virtual neighborhood and can only advance if it is a local
minimum, i.e., if
$\tau_{i}(t)\leq\min\{\tau_{i-1}(t),\tau_{i+1}(t),\tau_{r(i)}(t)\}$,
where $r(i)$ is the random connection of PE $i$. With probability
$(1-p)$ each  PE follows the original scheme. Note that the
occasional extra checking of the simulated time of the random
neighbor is not needed for the faithfulness of the
simulation. It is merely introduced to control the width of the time horizon.
Using a coarse-graining procedure analogous to
that used in \cite{KORN00_PRL} we find that the large-scale properties
of the virtual time horizon of our modified scheme are governed by the
equation,
%%%%%%%%%%%%%%%%%%% Equation (1): %%%%%%%%%%%%%%%%%%%%%%%%%%%%%%%%%%%%%
\begin{equation}
\frac{\partial \hat{\tau}}{\partial t} = -\gamma\hat{\tau} +
\frac{\partial^2 \hat{\tau}}{\partial x^2} - \lambda
\left(\frac{\partial \hat{\tau}}{\partial x}\right)^2 + \eta(x,t)
\;, \label{meaf_KPZ}
\end{equation}
%%%%%%%%%%%%%%%%%%%%%%%%%%%%%%%%%%%%%%%%%%%%%%%%%%%%%%%%%%%%%%%%%%%%%%%
where $\hat{\tau}(x,t)$ is the coarse-grained surface-height
fluctuation measured from the mean, and the temporal and spatial
derivatives are the coarse-grained interpretations of finite
differences. Similarly, $\eta(x,t)$ is a coarse-grained noise,
delta-correlated in space and time. The coefficients $\gamma$ and
$\lambda$ carry the details of the coarse-graining procedure. In
particular, $\gamma(p)$ is nonzero for all nonzero values of $p$
and $\gamma(p)\!\to\!0$ as $p\!\to\!0$, i.e., in this limit we
recover the KPZ equation for the original scheme. The
morphological properties of the surface (the virtual time horizon) governed by
Eq.~\ref{meaf_KPZ} are fundamentally different from the
one-dimensional KPZ surface for all $p\!\neq\!0$. The
algorithmic rules extended to include the quenched random
connections introduce a relaxational term (the first term
in Eq.~\ref{meaf_KPZ}) in the evolution of the virtual time
horizon. This term converts the system into a mean-field
growth model, whose time horizon is macroscopically
smooth, i.e., has a finite width.
To support the coarse-graining arguments, we simulated the exact
stochastic growth process defined by our modified conservative
algorithmic rules. Snapshots of the progress 
of the simulation, i.e., the virtual time horizon, are shown in
Fig.~1 for the original ($p\!=\!0.00$) and modified ($p\!=\!0.10$) cases.

The structure of the virtual time horizon can be most
transparently understood from the steady-state structure factor
$S(k)\!=\!\langle\tilde{\tau}_k\tilde{\tau}_{-k}\rangle/N_{\rm
PE}$, where $\tilde{\tau}_k$ is the spatial Fourier transform of
the surface fluctuations \cite{fourier_tr}. The presence of the
relaxational (first) term in the stochastic growth equation,
Eq.~\ref{meaf_KPZ}, implies that $\lim_{k\to 0}S(k)\!<\!\infty$, i.e.,
there are no large-amplitude long-wavelength modes in the surface.
Consequently, the width, 
$\langle w^2\rangle_{N_{\rm PE}}\!=\!(1/N_{\rm PE})\sum_{k\neq 0}S(k)$ 
is also finite. For further understanding we consider only the linear
terms in Eq.~\ref{meaf_KPZ} and obtain
%%%%%%%%%%%%%%%%%%%%%%% Equation(2): %%%%%%%%%%%%%%%%%%%%%%%%%%%%%%%%
\begin{equation}
S(k)\propto\frac{1}{\gamma + k^2}\;.
\label{S_k}
\end{equation}
%%%%%%%%%%%%%%%%%%%%%%%%%%%%%%%%%%%%%%%%%%%%%%%%%%%%%%%%%%%%%%%%%%%%%
In this approximation the lateral correlation length $\xi$ of the
surface fluctuations is $1/\sqrt{\gamma}$, i.e., it is finite
for all $p\!\neq\!0$. From Eq.~\ref{S_k} it also follows that the
local slopes of the virtual time horizon remain short-range
correlated and the utilization $\langle u(\infty)\rangle_{N_{\rm PE}}$
(average progress rate) is 
a nonzero constant in the limit $N_{\rm PE}\!\to\!\infty$, just
like in the original conservative scheme. 
The behavior of the structure factor $S(k)$ 
indicates (Fig.~2) that it approaches a finite value as
$k\!\to\!0$. Further, the linearized version of the theory seems to
work well in the small-$k$ regime (Fig.~2 inset), yielding a finite
correlation length ($\xi\!\approx\!26$ for $p\!=\!0.10$) 
for a long chain of PEs.

Choosing an infinitesimally small
$p$, the virtual time horizon becomes macroscopically smooth with
a finite width. At the same time, the utilization is only reduced
by an infinitesimal amount, as a result of the occasional extra
checking with the random neighbors.
This trade-off is substantially favorable for the
conservative PDES scheme: by giving up an infinitesimally small fraction of
the utilization, the width is reduced from infinity to a finite value,
making measurement and data management scalable under the
conservative PDES scheme. 
For example, for $p\!=\!0.10$, $\langle w^2(\infty)\rangle_{N_{\rm
PE}}$ is reduced from infinity to approximately
$5.25$ [Fig.~3(A)], while $\langle u(\infty)\rangle_{N_{\rm PE}}$ 
is reduced by only about $1.6\%$ [Fig.~3(B)].
One can also observe the clear self-averaging property for both global
observables, the width and the utilization (Fig.~3).
From a broader statistical physics viewpoint, one can ask
whether a small, non-zero fraction $q$ of the PEs with random links
(checked, e.g., at every step) is sufficient to control the width.
While this choice clearly weakens load balancing and the 
utilization [Fig.~3(B)], our results in Fig.~3(A) and recent work
on the closely related XY-model on a small-world network \cite{XY_sw}
suggest that a finite width is achieved.  
There is growing evidence that systems without inherent frustration
exhibit mean-field characteristics when the original short-range
interaction topology is modified to a small-world network
\cite{NEWMAN,BARRAT,GITTERMAN,XY_sw}.

The generalization when random links are added to a higher-dimensional
underlying regular lattice is clear: since the one-dimensional
case with random links is governed by the mean-field equation, in
higher dimensions it will be even more so (i.e., the
critical dimension \cite{BARA95} of the model with random links is
less than one).
The generalization for the many-sites-per-PE case also follows from
universality arguments: without the random connections, 
there is an additional fast-roughening phase for early times when the
evolution of the time horizon corresponds to random deposition
\cite{BARA95}. Subsequently, it will cross 
over to the KPZ growth regime and finally saturate. With many sites
per PE, the typically desired and efficient way of implementing PDES
\cite{KORN99_JCP,LUBA88}, the saturation value of the width can become
extremely large. This underscores the importance of implementing the additional
synchronizations through random links to suppress the roughness of the
time horizon.

%%%%%%%%%%%%%%%%%%%%%%%%%%%% REFERENCES %%%%%%%%%%%%%%%%%%%%%%%%%%%%%%%%

\bibliography{scibib}

\begin{thebibliography}{99}
\bibitem{fourier_tr}
Here we define the Fourier transform of the surface fluctuations as 
%%%%%%%%%%%%%%%%%%%%%%%%%%%%%%%%%%%%%%%%%%%%%%%%%%%%%%%%%%%%%%%%%%%
$\tilde{\tau}_k=\sum_{j=1}^{N_{\rm PE}}e^{-ikj}(\tau_j-\overline\tau)$, 
%%%%%%%%%%%%%%%%%%%%%%%%%%%%%%%%%%%%%%%%%%%%%%%%%%%%%%%%%%%%%%%%%%%
where $k=2\pi n/N_{\rm PE}$ and $n=0,1,2,\ldots,N_{\rm PE}\!-\!1$.
\end{thebibliography}

\bibliographystyle{Science}

\clearpage

%%%%%%%%%%%%%%%%%%%%%% FIGURES %%%%%%%%%%%%%%%%%%%%%%%%%%%%%%%%%

%%%%%%%%%%%%%%%%%%%%%%%% FIG1 %%%%%%%%%%%%%%%%%%%%%%%%%%%%%%%%%%%%%%%%%
\begin{figure}
\vspace*{3.0truecm}
\includegraphics{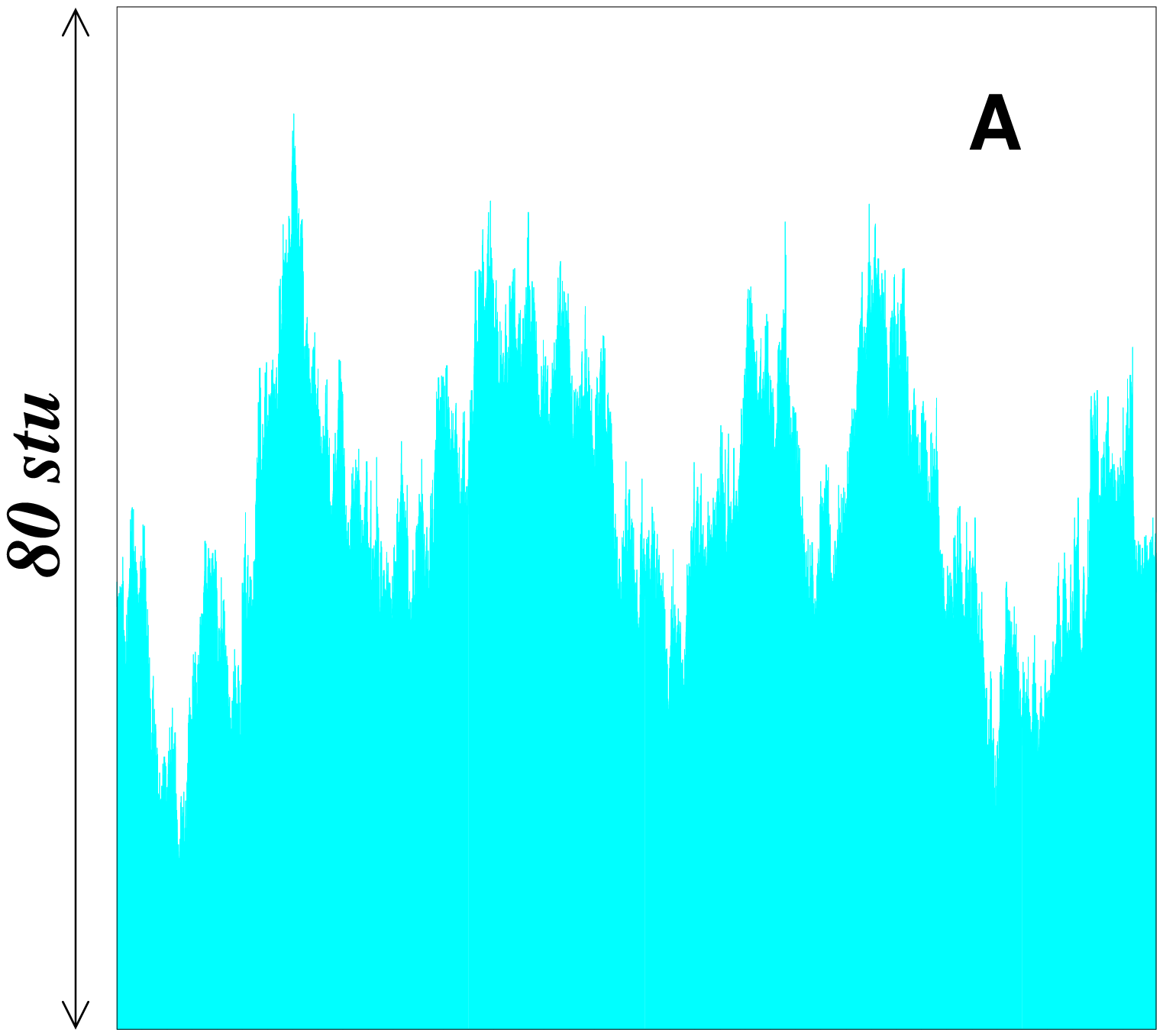}
\includegraphics{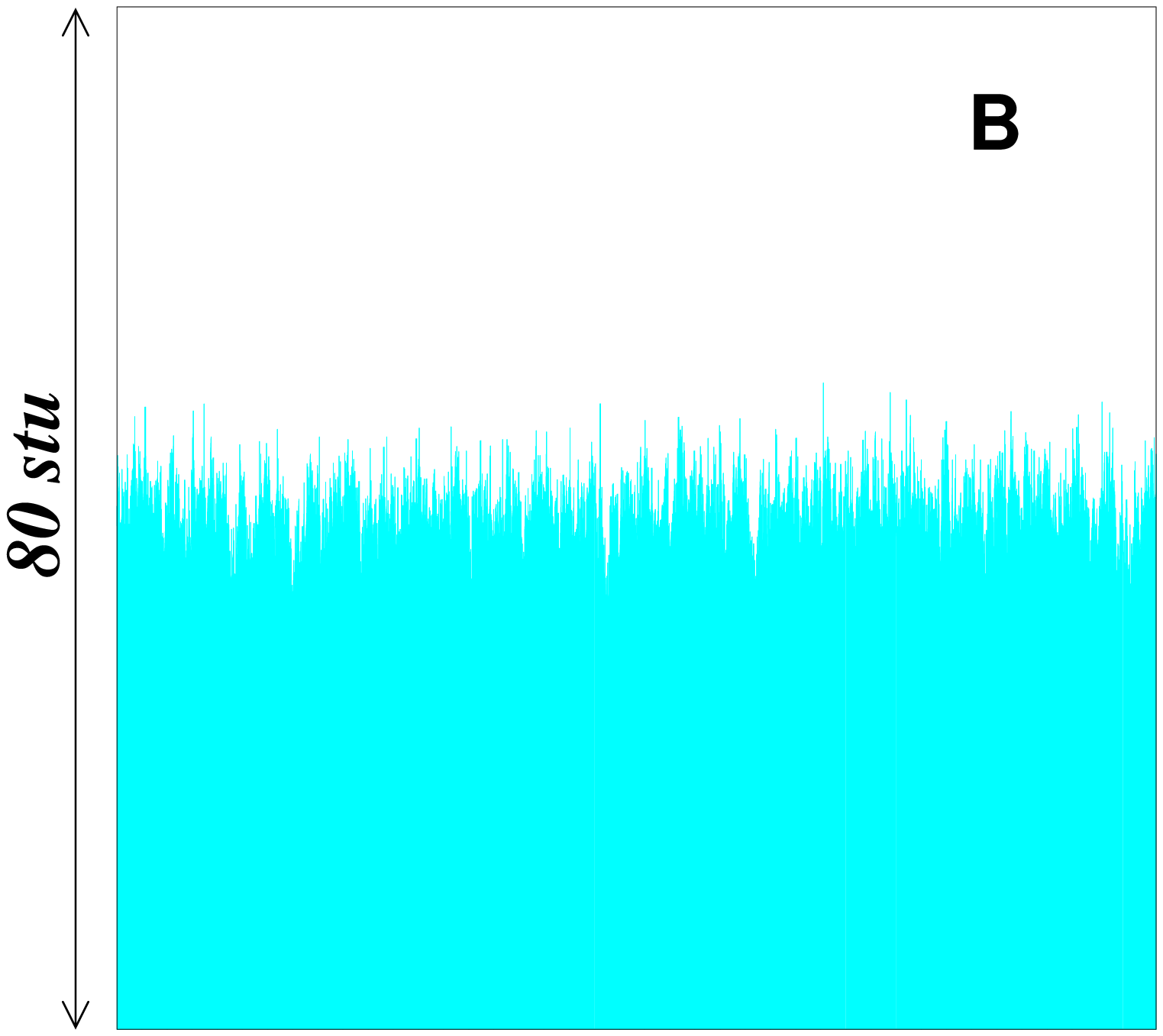}
\vspace*{1.0truecm}
\caption{Snapshots of the virtual time
horizons in the steady state for the one-site-per PE conservative
PDES scheme with $N_{\rm PE}\!=\!10,000$ after $t\!=\!10^6$
parallel algorithmic steps for ({\bf A}) the original algorithm
($p\!=\!0.00$) and ({\bf B}) the modified one ($p\!=\!0.10$) with quenched
random connections. The vertical scale, $80$ simulated time units
(stu) is the same in ({\bf A}) and ({\bf B}).}
\label{fig1}
\end{figure}
%%%%%%%%%%%%%%%%%%%%%%%%%%%%%%%%%%%%%%%%%%%%%%%%%%%%%%%%%%%%%%%%%%%%%%%

%%%%%%%%%%%%%%%%%%%%%%%%% FIG2 %%%%%%%%%%%%%%%%%%%%%%%%%%%%%%%%%%%%%%%%
\begin{figure}
\vspace*{4.0truecm}
\includegraphics{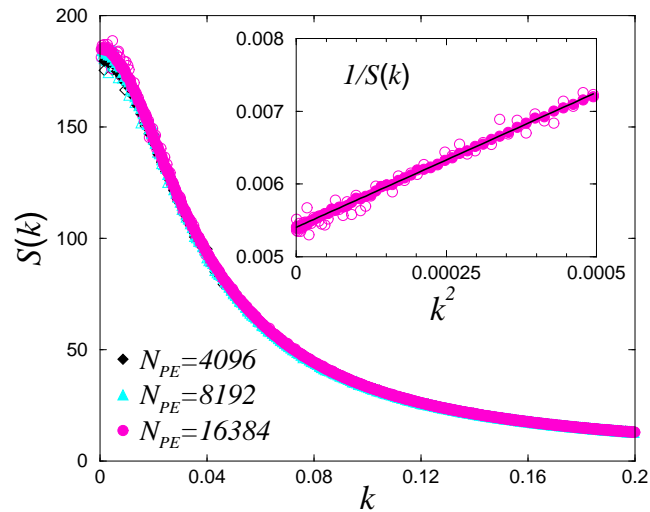}
\vspace*{3.0truecm}
\caption{Steady-state structure factor of the virtual time
horizon for $p\!=\!0.10$. In addition to ensemble averages over $100$
realizations of the random links (filled symbols), single realizations
(the same open symbols) are also shown. The inset shows a magnified
view of $1/S(k)$ versus $k^2$ for small $k$ for the largest
system. The straight line is the best linear fit used to determine the
correlation length.}
\label{fig2}
\end{figure}
%%%%%%%%%%%%%%%%%%%%%%%%%%%%%%%%%%%%%%%%%%%%%%%%%%%%%%%%%%%%%%%%%%%%%%%

%%%%%%%%%%%%%%%%%%%% FIG3 %%%%%%%%%%%%%%%%%%%%%%%%%%%%%%%%%%%%%%%%%%%%%
\begin{figure}
\vspace*{2.8truecm}
\includegraphics{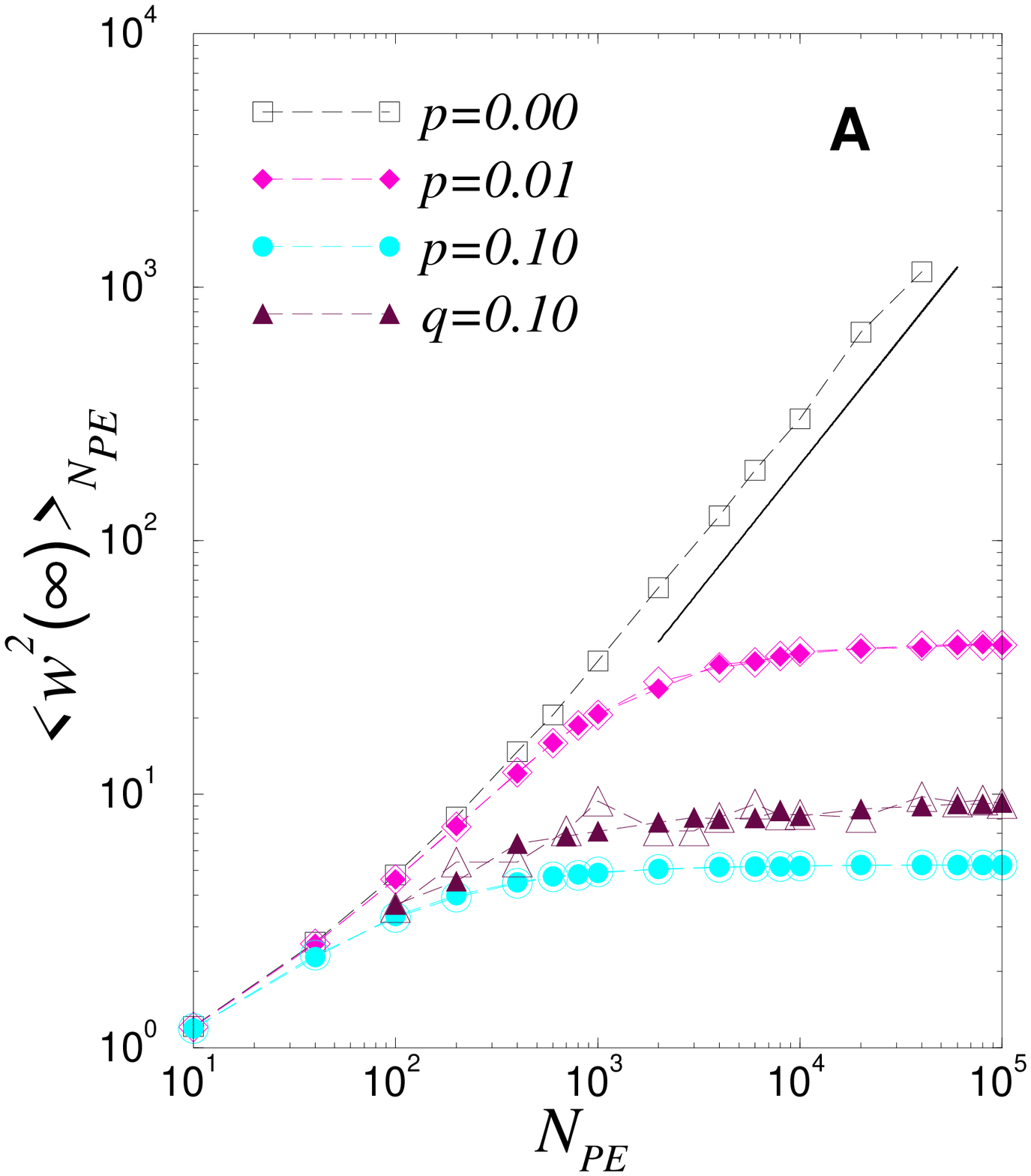}
\includegraphics{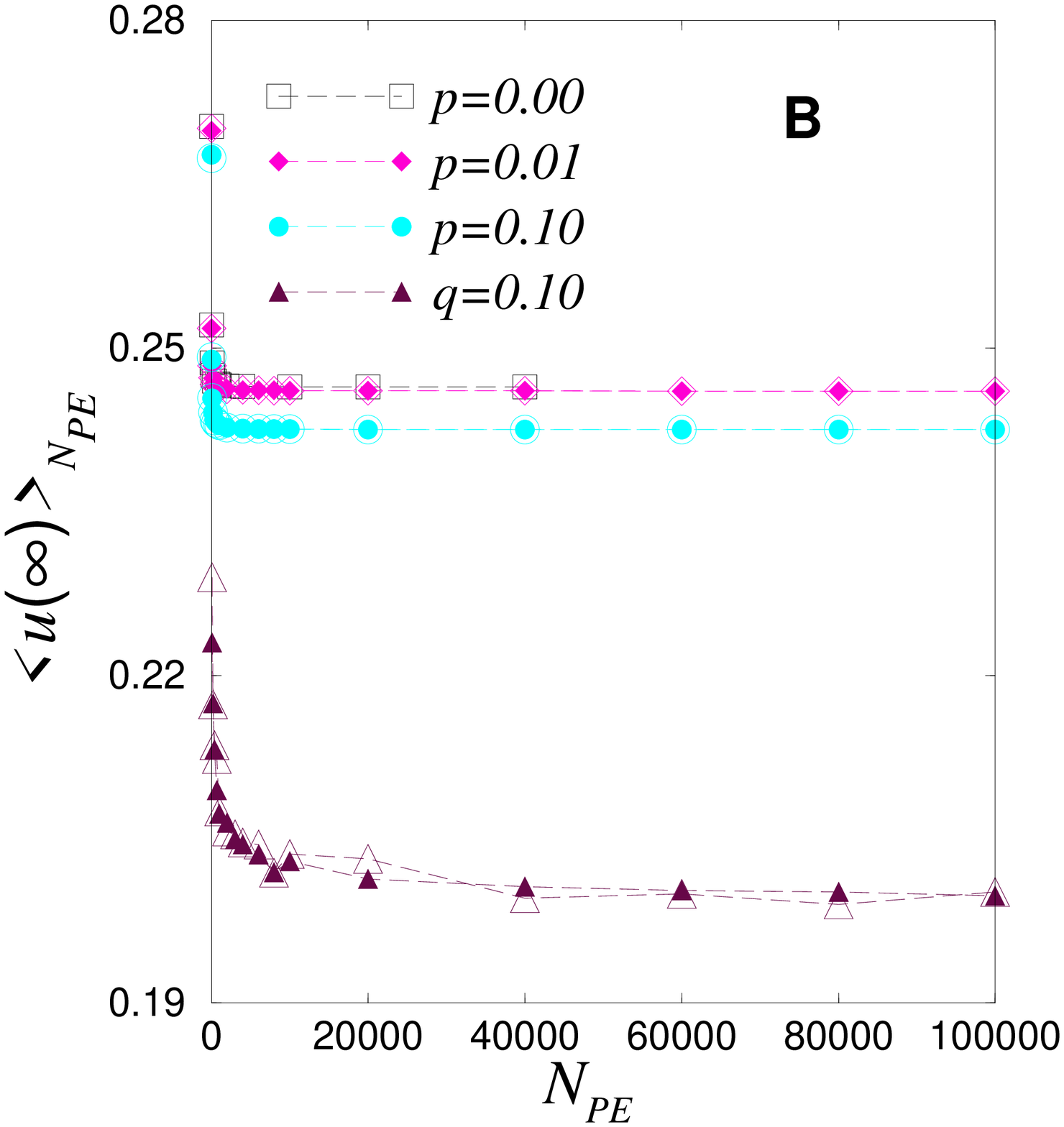}
\vspace*{2.0truecm}
\caption{({\bf A}) Average steady-state width of the virtual time
horizon for various values of $p$ for the one-site-per-PE PDES scheme.
In addition to ensemble averages over $10$ realizations of the random
links (filled symbols), a single realization is also shown (the same
open symbols). The solid straight line represents the asymptotic
one-dimensional KPZ power-law divergence with roughness exponent
$\alpha\!=\!1/2$ for the $p\!=\!0$ case. Note the log-log scales. 
The $q\!=\!0.10$ data set corresponds to the case when only 10\% of the PEs
have random links and those are checked at every step. 
({\bf B}) The steady-state utilization (fraction of non-idling PEs) for the
same cases as in ({\bf A}).}
\label{fig3}
\end{figure}
%%%%%%%%%%%%%%%%%%%%%%%%%%%%%%%%%%%%%%%%%%%%%%%%%%%%%%%%%%%%%%%%%%%%%%%

\end{document}